\documentstyle[12pt,epsf]{article}

\def\siml{{\
\lower-1.2pt\vbox{\hbox{\rlap{$<$}\lower6pt\vbox{\hbox{$\sim$}}}}\ }}  
\def\bfnabla{\mbox{\boldmath $\nabla$}}

\def\bfsigma{\mbox{\boldmath $\sigma$}}
\def\bftau{\mbox{\boldmath $\tau$}}

\def\al{\alpha}

\newcommand{\nn}{\nonumber}
\newcommand{\be}{\begin{equation}}
\newcommand{\ee}{\end{equation}}
\newcommand{\bea}{\begin{eqnarray}}
\newcommand{\eea}{\end{eqnarray}}

\def\dsl{\,\raise.15ex\hbox{/}\mkern-13.5mu D}

\newcommand{\Appendix}[1]%
    {%
     \section{#1}%
      }

\begin{document}\setlength{\unitlength}{1mm}

\begin{titlepage}
\begin{flushright}
\tt{UB-ECM-PF 02/26}
\end{flushright}

\vspace{1cm}
\begin{center}
\begin{Large}
{\bf Leading Chiral Logarithms to the Hyperfine Splitting of the Hydrogen
and Muonic Hydrogen}\\[2cm] 
\end{Large} 
{\large Antonio Pineda}\footnote{pineda@ecm.ub.es}\\
{\it Dept. d'Estructura i Constituents de la Mat\`eria and IFAE,
U. Barcelona \\ Diagonal 647, E-08028 Barcelona, Catalonia, Spain}
\end{center}

\vspace{1cm}

\begin{abstract}
We study the hydrogen and muonic hydrogen within an effective field
theory framework. We perform the matching between heavy baryon
effective theory coupled to photons and leptons and the relevant
effective field theory at atomic scales. This matching can be
performed in a perturbative expansion in $\al$, $1/m_p$ and the chiral
counting. We then compute the $O(m_{l_i}^3\al^5/m_p^2\times logarithms)$
contribution (including the leading chiral logarithms) to the Hyperfine
splitting and compare with experiment. They can explain about 2/3 of
the difference between experiment and the pure QED prediction when
setting the renormalization scale at the $\rho$ mass. We give an
estimate of the matching coefficient of the spin-dependent
proton-lepton operator in heavy baryon effective theory.
\vspace{5mm} \\
PACS numbers: 12.39.Fe, 11.10.St, 12.39.Hg, 12.20.Ds
\end{abstract}

\end{titlepage}
\vfill
\setcounter{footnote}{0} 
\vspace{1cm}

\section{Introduction}
Years have passed since the advent of QCD. After numerous attempts to
understand QCD by using several models, more studies now move towards
trying to parameterize the QCD properties in a model independent way
with the help of different systematics that are usually highlighted
by the specific kinematic situation under study. One could hope that
this approach may bring some light in the understanding of QCD or a
least to provide some consistency check between different
models. Therefore, it becomes important to be able to relate as many
observables as possible in a model independent framework. Effective
field theories (EFT's) may play an important role in this approach.

Within the above philosophy, the study of hydrogen ($ep$) and muonic
hydrogen ($\mu p$), in particular of the high precision measurement of
different splittings, can provide accurate determinations of some
hadronic parameters related with the proton elastic and inelastic
electromagnetic form factor, like the proton radius and magnetic
moment, polarization effects, etc....

In the $ep$ and $\mu p$ we are basically testing the proton with
different probes ($e$, $\mu$, $\gamma$). They correspond to the
simplest possible probes since they are point-like particles and the
interaction is perturbative (the analogy with deep inelastic
scattering is evident and it has already been used since long ago
\cite{Zemach,Iddings,DS,DeRafael} in order to obtain some of these
hadronic parameters from dispersion relations). They also provide the
first natural step towards more complicated systems like exotic or
heavy (muonic) atoms.

\medskip

The $ep$ and $\mu p$ systems are, in a first approximation, states
weakly bound by the Coulomb interaction and their typical binding
energy and relative momentum are, $E\sim m_{e(\mu)}\al^2$ and $|{\bf
p}| \sim m_{e(\mu)}\al$, respectively. We will switch off the weak
interactions in this work. Therefore, the $ep$ and $\mu p$ systems
become stable and $C$, $P$ and $T$ are exact symmetries of these
systems. In any case, several different scales are involved in their
dynamics:

For the $e p$ system they are: $\dots$, $m_{e}\al^2$ , $m_{e}\al$,
$m_{e}$, $\Delta m=m_n-m_p$, $\Delta=m_{\Delta}-m_p$, $m_{\pi}$,
$m_p$, $m_\rho$, $\Lambda_{\chi}$, $\dots$ that we will
group and name in the following way:
\begin{itemize}
\item
$m_{e}\al^2$: ultrasoft (US) scale.
\item
$m_{e}\al$: soft scale.
\item
$\mu_{e p}= {m_{e}m_p \over m_{e} +m_p}$, $\Delta m=m_n-m_p$, $m_e$:
hard scale. 
\item
$m_{\mu}$, $\Delta=m_{\Delta}-m_p$, $m_{\pi}$: pion scale.
\item
$m_p$, $m_\rho$, $\Lambda_{\chi}$: chiral scale.
\end{itemize}

For the $\mu p$ system they are: $\dots$, $m_{\mu}\al^2$ , $m_{\mu}\al$,
$m_{\mu}$, $\Delta m=m_n-m_p$, $m_e$, $\Delta=m_{\Delta}-m_p$,
$m_{\pi}$, $m_p$, $m_\rho$, $\Lambda_{\chi}$, $\dots$ that we
will group and name in the following way:
\begin{itemize}
\item
$m_{\mu}\al^2$: US scale.
\item
$\Delta m=m_n-m_p$, $m_e$, $m_{\mu}\al$: soft scale.
\item
$\mu_{\mu p}= {m_{\mu}m_p \over m_{\mu} +m_p}$, $m_{\mu}$,
$\Delta=m_{\Delta}-m_p$, $m_{\pi}$: hard/pion scale.
\item
$m_p$, $m_\rho$, $\Lambda_{\chi}$: chiral scale.
\end{itemize}

By doing ratios with the different scales, several small expansion
parameters can be built. Basically, this will mean that the
observables, the spectrum in our case, can be written, up to large
logarithms, as an expansion, in the case of the $ep$, in $\al$, ${m_e \over
m_{\pi}}$ and ${m_{\pi} \over m_p}$, and in the case of the $\mu$$p$,
in $\al$ and ${m_\mu \over m_p}$. It will also prove convenient
sometimes to use the reduced mass $\mu_{\mu(e) p}$, since it will
allow to keep (some of) the exact mass dependence at each order in
$\al$. In order to be more precise, the $e p$ energy will be expanded
in the following way (up to logarithms):
\be
E(ep)=-{\mu_{e p}\al^2 \over 2n^2}(1+c_2\al^2+c_3\al^3+\cdots)\,,
\ee
where 
\be
c_n=\sum_{i,j=0}^{\infty}c_n^{(i,j)}
\left(
{m_e \over m_\pi}
\right)^i
\left(
{m_\pi \over m_p}
\right)^j
 + \cdots
\,,
\ee
and $c_n^{(i,j)}$ are functions of dimensionless quantities of $O(1)$ like 
${\mu_{e p}/m_e}$, $m_\mu/m_\pi$, etc. 

For the $\mu p$ things work analogously: 
\be
E(\mu p)=-{\mu_{\mu p}\al^2 \over 2n^2}(1+c_1\al+c_2\al^2+c_3\al^3 +\cdots)
\,,
\ee
where $c_1$ does not depend on hadronic quantities, only on $m_\mu\al/m_e$, 
and $(n \geq 2)$
\be
c_n=\sum_{j=0}^{\infty}c_n^{(j)}
\left(
{m_\pi \over m_p}
\right)^j
 + \cdots
\,,
\ee
where $c_n^{(j)}$ are functions of dimensionless quantities of $O(1)$ like 
${\mu_{\mu p}/m_\mu}$, $m_\mu/m_\pi$, $m_\mu\al/m_e$, etc. 

Let us stress that the coefficients $c_n$ can be expanded in the ratio
  ${m_{\pi} \over m_p}$, i.e. in the chiral/heavy-baryon expansion
  ($m_p$ should also be understood as $\Lambda_{\chi}$).

\medskip

In order to disentangle all the different scales mentioned above it is
convenient to use EFT's. In order to obtain the relevant one for these
systems we first need to decide what are the degrees of freedom we
want to describe. In our case we want to describe the $ep$ and $\mu p$
systems at ultrasoft or smaller energies. Therefore, degrees of
freedom with higher energies can (and will) be integrated out in order
to obtain the EFT to describe these systems. One EFT that fulfills
this requirement is potential NRQED (pNRQED) \cite{Mont,pos} (for some
applications see \cite{KP} and see also
\cite{pos6,Yelkhovsky}). pNRQED appears after integrating out the soft
scale from NRQED \cite{NRQED} and it shares some similarities with the
approach followed in Ref. \cite{Pachucki4}. We will obtain pNRQED by
passing through different intermediate effective field theories after
integrating out different degrees of freedom. The path that we will
take is the following (in some cases, instead of this chain of EFT's
one can use dispersion relations, or direct experimental data, in
order to obtain the matching coefficients):
$$
{\rm HBET} \rightarrow ({\rm QED}) \rightarrow {\rm NRQED} \rightarrow
{\rm pNRQED}.
$$
This way of working opens the possibility to compute the observables of
atomic physics with the parameters obtained from heavy baryon
effective theory (HBET), which is much close to QCD since it
incorporates its symmetries automatically, in particular the chiral
symmetry. Besides, it is the matching with HBET that will allow to
relate the matching coefficients used for $ep$ with the ones used in
$\mu p$.  HBET \cite{HBET} describes systems with one heavy baryon:
the proton, the neutron or also the delta \cite{DHBET} at the pion
mass scale. The chiral scale explicitly appears in the Lagrangian as an
expansion in $1/\Lambda_{\chi}$, $1/m_p$ and any other smaller scale
remains dynamical in this effective theory. In short, HBET is a EFT
defined with an UV cut-off $\nu$ such that $\nu \ll \Lambda_{\chi}$
but larger than any other dynamical scale in the problem.

In the $\mu p$, NRQED appears after integrating out the hard scale,
whereas in the $e p$, NRQED appears after integrating out the pion and
hard scales. In this last case one could pass through an intermediate
theory (QED) defined by integrating out the pion scale and profit from
the fact that pion and hard scales are widely separated. Nevertheless,
we will do the matching here in one step for simplicity.

pNRQED is obtained after integrating out the soft scale. We refer to
\cite{Mont,pos} for further details.

The above methodology allows to compute (or parameterize in a model
independent way) the coefficients $c$'s in a systematic expansion in
the different small parameters on which these systems depend.

It is the aim of this paper to use this procedure for the computation
of the leading-logarithm hadronic contributions to the hyperfine splitting
for both the $ep$ and $\mu p$. This means to compute the
spin-dependent piece of $c_3$ with $O((m_e/m_p)^2\times logarithms)$ and
$O((m_\mu/m_p)^2\times logarithms)$ accuracy for the $ep$ and $\mu p$
respectively.

\section{Effective Field Theories}

In this section, we will consider the different EFTs that will be
necessary for our calculation.

\subsection{HBEFT}
\label{secHBET}

Our starting point is the SU(2) version of HBEFT coupled to leptons,
where the delta is kept as an explicit degree of freedom.  The degrees
of freedom of this theory are the proton, neutron and delta, for which
the non-relativistic approximation can be taken, and pions, leptons
(muons and electrons) and photons, which will be taken relativistic.

Our first aim will be to present the effective Lagrangian of this
theory. It corresponds to a hard cut-off $\mu << m_p$,
$\Lambda_{\chi}$ and much larger than any other scale in the problem.
The Lagrangian can be split in several sectors. Most of them have
already been extensively studied in the literature but some will be
new.  Moreover, the fact that some particles will only enter through
loops, since only some specific final states are wanted, will simplify
the problem. The Lagrangian can be structured as follows
\be
{\cal L}_{HBET}=
{\cal L}_{\gamma}
+
{\cal L}_{l}
+
{\cal L}_{\pi}
+
{\cal L}_{l\pi}
+
{\cal L}_{(N,\Delta)}
+
{\cal L}_{(N,\Delta)l}
+
{\cal L}_{(N,\Delta)\pi}
+
{\cal L}_{(N,\Delta)l\pi},
\ee
representing the different sectors of the theory. In particular, the
$\Delta$ stands for the spin 3/2 baryon multiplet (we also use
$\Delta=m_{\Delta}-m_p$, the specific meaning in each case should be
clear from the context).

The Lagrangian can be written as an expansion in $e$ and $1/m_p$. Our
aim is to obtain the hyperfine splitting with
$O(m_{l_i}^3\al^5/m_p^2\times(\ln m_q,\ln \Delta,\ln m_{l_i}))$
accuracy, where $m_q$ stands for the mass of the light, $u$ and $d$
(or $s$), quarks and $m_{l_i}$ for the mass of the lepton (the leading
order contribution to the hyperfine splitting reads $E_F
=(8/3)c_F^{(p)}m_{l_i}^2\al^4/m_p$, where $c_F^{(p)}$ is defined in
Eq. (\ref{LNdelta})). Therefore, we need,
in principle, the Lagrangian with $O(1/m_p^2)$ accuracy. Let us
consider the different pieces of the Lagrangian more in detail.

The photonic Lagrangian reads (the first corrections to this term
scale like $\al^2/m_p^4$)
\be
\label{Lg}
{\cal L}_\gamma=-\frac{1}{4}F^{\mu\nu}F_{\mu \nu} 
\,.
\ee
The leptonic sector reads ($i D_\mu=i\partial_\mu-eA_\mu$)
\be
\label{Ll}
{\cal L}_l=\sum_i \bar l_i  (i\dsl-m_{l_i}) l_i
\,,
\ee
where $i=e,\mu$. We do not include the term 
\be
-{e g_{l_i}
\over m_p}\bar l_i \sigma_{\mu\nu}l_iF^{\mu\nu}
\,,
\ee
since the coefficient $g_{l_i}$ is suppressed by powers of $\al$ and 
the mass of the lepton. Therefore, it would give contributions beyond the
accuracy we aim. In any case, any eventual contribution would be
absorbed in a low energy constant.

The pionic Lagrangian ${\cal L}_{\pi}$ is usually organized in the
chiral counting. From the analysis of the next section we will see
that the free pion propagators provide with the necessary precision.
Therefore, we only need the free-particle pionic Lagrangian:
\be
{\cal L}_\pi=(\partial_\mu \pi^+)(\partial^\mu \pi^-)-m_\pi^2\pi^+\pi^-
+{1 \over 2}(\partial_\mu \pi^0)(\partial^\mu \pi^0)-{1 \over
2}m_\pi^2\pi^0\pi^0 
\,.
\ee

The one-baryon Lagrangian ${\cal L}_{(N,\Delta)\pi}$ is needed at
$O(1/m_p^2)$.  Nevertheless a closer inspection simplifies the
problem. A chiral loop produces a factor $1/(4\pi F_0)^2 \sim
1/m_p^2$.  Therefore, the pion-baryon interactions are only needed at
$O(m_\pi)$, the leading order, which is known
\cite{HBET,DHBET}.\footnote{Actually terms that go into the physical
mass of the proton and into the physical value of the anomalous
magnetic moment of the proton $\mu_p=c_F^{(p)}-1$ should also be
included (at least in the pure QED computations) and it will be 
assumed in what follows. For our computation these effects would be 
formally subleading. In any case, their
role is just to bring the {\it bare} values of $m_0$ and $\mu_0$ to
their physical values. Therefore, once the values of $m_p$ and $\mu_p$
are measured by different experiments, they can be distinguished from
the effects we are considering in this paper.} For the explicit
expressions we refer to these references.

Therefore, we only need the one-baryon Lagrangian ${\cal
L}_{(N,\Delta)}$ at $O(1/m_p^2)$ coupled to electromagnetism. This
would be a NRQED-like Lagrangian for the proton, neutron (of spin 1/2)
and the delta (of spin 3/2). The neutron is actually not needed at
this stage. The relevant term for the proton reads
\be
\label{LNdelta}
\delta {\cal L}_{(N,\Delta)}= N^\dagger_{p} \Biggl\{iD_0+ {{\bf
D}^2_p\over 2 m_{p}} + 
 {{\bf D}_p^4\over
8 m^3_{p}} - eZ_p {c_F^{(p)} \over 2m_p}\, {\bf \bfsigma \cdot B} 
 - ie Z_p { c_S^{(p)} \over 8m_p^2}\, \bfsigma \cdot \left({\bf D}_p
     \times {\bf E} -{\bf E}    \times {\bf D}_p\right) 
\Biggr\} N_{p}
\,,
\ee
where $ i D^0_p=i\partial_0 +Z_peA^0$, $i{\bf D}_p=i{\bfnabla}-Z_pe{\bf A}$. 
For the proton $Z_p=1$. We have not included a term like
\be
{c_D^{(p)} \over m_p^2}
  N^\dagger_{p} \left[{\bf \bfnabla \cdot E }\right]N_{p}
\,.
\ee
We could have done so but it may also be eliminated by some field
redefinitions. In any case it would give contribution to the
spin-independent terms so we will not consider it further in this
work.
 
As for the delta (of spin 3/2), it mixes with the nucleons at
$O(1/m_p)$ ($O(1/m_p^2)$ are not needed in our case). The only
relevant interaction in our case is the $p$-$\Delta^+$-$\gamma$ term,
which is encoded in the second term of 
\be
\delta {\cal L}_{(N,\Delta)}
= 
T^{\dagger}(i\partial_0-\Delta)T
+{eb_{1,F} \over 2m_p}
\left(
T^{\dagger}\bfsigma ^{(3/2)}_{(1/2)}\cdot {\bf
B}\,\bftau^{3(3/2)}_{(1/2)} N + h.c.
\right)
\,,
\ee  
where $T$ stands for the delta 3/2 isospin multiplet, $N$ for the nucleon 1/2
isospin multiplet and the transition spin/isospin matrix elements fulfill
(see \cite{Weise})
\be
\bfsigma^{i(1/2)}_{(3/2)}\bfsigma^{j(3/2)}_{(1/2)}
={1 \over 3}(2\delta^{ij}-i\epsilon^{ijk}\bfsigma^k),
\qquad
\bftau^{a(1/2)}_{(3/2)}\bftau^{b(3/2)}_{(1/2)}
={1 \over 3}(2\delta^{ab}-i\epsilon^{abc}\bftau^c).
\ee

The baryon-lepton Lagrangian provides new terms that are not usually considered 
in HBET. The relevant term in our case is the interaction between 
the leptons and the nucleons (actually only the proton):
\be
\label{LNl}
\delta {\cal L}_{(N,\Delta)l}=\displaystyle\frac{1}{m_p^2}\sum_i c_{3,R}^{pl_i}
{\bar N}_p \gamma^0
  N_p \ \bar{l}_i\gamma_0 l_i
+\displaystyle\frac{1}{m_p^2}\sum_i c_{4,R}^{pl_i}{\bar N}_p \gamma^j \gamma_5
N_p 
  \ \bar{l}_i\gamma_j \gamma_5 l_i
\,.
\ee
The above matching coefficients fulfill 
$c_{3,R}^{pl_i}=c_{3,R}^{p}$ and $c_{4,R}^{pl_i}=c_{4,R}^{p}$ up to terms
suppressed by $m_{l_i}/m_p$, which will be sufficient for our purposes.

Let us note that with the conventions above, $N_p$ is the field of the proton 
(understood as a particle) with
positive charge if $l_i$ represents the leptons (understood as
particles) with negative charge.

This finishes all the needed terms for this paper, since the other sectors
of the Lagrangian would give subleading contributions.

\subsection{NRQED($\mu$)}
\label{NRQED(mu)}

In the muon-proton sector, by integrating out the $m_\pi$ scale, an
effective field theory for muons, protons and photons appears. In
principle, we should also consider neutrons but they play no role at
the precision we aim. The effective theory corresponds to a hard
cut-off $\nu << m_\pi$ and therefore pions and deltas have been
integrated out. The Lagrangian is equal to the previous case but
without pions and deltas and with the following modifications: ${\cal
L}_{l} \rightarrow {\cal L}_{e}+{\cal L}^{(NR)}_\mu$ and ${\cal
L}_{(N,\Delta)l} \rightarrow {\cal L}_{Ne}+{\cal L}^{(NR)}_{N\mu}$,
where it is made explicit that the the muon has become no
relativistic. Any further difference goes into the matching
coefficients, in particular, into the matching coefficients of the
baryon-lepton operators. In summary, the Lagrangian reads
\be
{\cal L}_{NRQED(\mu)}=
{\cal L}_{\gamma}
+
{\cal L}_{e}
+
{\cal L}^{(NR)}_{\mu}
+
{\cal L}_{N}
+
{\cal L}_{Ne}
+
{\cal L}_{N\mu}^{(NR)}
\,,
\ee
where 
\be
{\cal L}^{(NR)}_{\mu}= l^\dagger_{\mu} \Biggl\{ i D_{\mu}^0 + \, {{\bf D}_{\mu}^2\over 2 m_{\mu}} +
 {{\bf D}_{\mu}^4\over
8 m^3_{\mu}} + eZ_{\mu} {c_F^{(\mu)} \over 2m_\mu}\, {\bf \bfsigma \cdot B}
 +
 i eZ_{\mu} {c_S^{(\mu)} \over 8 m^2_{\mu}}\,  {\bf \bfsigma \cdot \left(D_{\mu} \times
     E -E \times D_{\mu}\right) } 
\Biggr\} l_{\mu}
\ee
and
\be
{\cal L}_{N\mu}^{NR}=
\displaystyle\frac{c_{3,NR}^{pl_\mu}}{m_p^2} N_p^{\dagger}
  N_p \ {l}^{\dagger}_\mu l_\mu
-\displaystyle\frac{c_{4,NR}^{pl_\mu}}{m_p^2} N_p^{\dagger}{\bfsigma}
  N_p \ {l}^{\dagger}_\mu{\bfsigma} l_\mu
\,,
\ee
with the following definitions: $ i D^0_\mu=i\partial_0 -Z_\mu eA^0$,
$i{\bf D}_\mu=i{\bfnabla}+Z_\mu e{\bf A}$ and $Z_\mu=1$. ${\cal L}_e$
stands for the relativistic leptonic Lagrangian (\ref{Ll}) and ${\cal
L}_{Ne}$ for Eq. (\ref{LNl}), both for the electron case only.

A term of the type
\be
-{e g_{l_e}
\over m_\mu}\bar l_e \sigma_{\mu\nu}l_eF^{\mu\nu}
\ee
is not taken into account because of the same reason as in Sec. \ref{secHBET}.
 
\subsection{QED($e$)}

After integrating out scales of $O(m_\pi)$ in the electron-proton
sector, an effective field theory for electrons coupled to protons
(and photons) appears.  Again, we should also consider neutrons but
they play no role at the precision we aim. This effective theory has a
cut-off $\nu << m_\pi$ and pions, deltas and muons have been
integrated out. The Lagrangian reads
\be
{\cal L}_{QED(e)}=
{\cal L}_{\gamma}
+
{\cal L}_{e}
+
{\cal L}_{N}
+
{\cal L}_{Ne}
\,.
\ee
This Lagrangian is similar to the previous subsection but without the muon.

\subsection{NRQED($e$)} 

After integrating out scales of $O(m_e)$ in the electron-proton
sector, we still have an effective field theory for electrons coupled
to protons and photons.  Nevertheless, now, the electrons are
non-relativistic. The Lagrangian is quite similar to the one in
subsec. \ref{NRQED(mu)} but without a light fermion and with the
replacement $\mu \rightarrow e$. The Lagrangian reads
\be
{\cal L}_{NRQED(e)}=
{\cal L}_{\gamma}
+
{\cal L}^{(NR)}_{e}
+
{\cal L}_{N}
+
{\cal L}_{Ne}^{(NR)}
\,.
\ee

\subsection{pNRQED}

After integrating out scales of $O(m_{l_i}\al)$ one ends up in a
Schr\"odinger-like formulation of the bound-state problem. We refer to
\cite{Mont,pos} for details.  The pNRQED Lagrangian for the $ep$ (the
non-equal mass case) can be found in Appendix B of the second Ref. in
\cite{pos} up to $O(m\al^5)$. The
pNRQED Lagrangian for the $\mu p$ is similar except for the fact that
light fermion (electron) effects have to be taken into account. The
explicit Lagrangian and a more detailed analysis of this case will be
presented elsewhere. For the purposes of this paper, we only have to
consider the spin-dependent delta potential:
\be
\delta V = 2{c_{4,NR} \over m_p^2}{\bf S}^2\delta^{(3)}({\bf r})
\,.
\ee
which will contribute to the hyperfine splitting.

\section{Form Factors: definitions}

It will turn out convenient to introduce some notation 
before performing the matching between HBET and NRQED. 
We first define the form factors, which we will understand 
as pure hadronic quantities, i.e. without electromagnetic 
corrections. 

Our notation is based on the 
one of Ref. \cite{Pantforder}. We define
$J^\mu=\sum_i Q_i{\bar q}_i\gamma^\mu q_i$ where $i=u,d$ (or $s$). The 
form factors are then defined by the following equation:   
\be
\langle {p^\prime,s}|J^\mu|{p,s}\rangle
=
\bar u(p^\prime) \left[ F_1(q^2) \gamma^\mu +
i F_2(q^2){\sigma^{\mu \nu} q_\nu\over 2 m} \right]u(p)
\label{current}
\,,
\ee
where $q=p'-p$  and $F_1$, $F_2$ are the Dirac and Pauli form factors,
respectively. The states
are normalized in the following (standard relativistic) way:
\be
\langle p',\lambda'|p,\lambda \rangle
=
(2\pi)^3 2p^0\delta^3({\bf p}'-{\bf p})
\delta_{\lambda' \lambda}\,,
\ee
and 
\be
u(p,s){\bar u}(p,s)={\rlap/P+m_p \over 2m_p}{1+\gamma_5{\rlap/s} \over 2}
\,,
\ee
where $s$ is an arbitrary spin four vector obeying $s^2=-1$ and
$P\cdot s=0$.
 
The form factors could be (analytically) expanded as follows 
\be
F_i(q^2)=F_i+{q^2 \over m^2}F_i^{\prime}+...\,
\ee 
for very low momentum. Nevertheless, we will be interested instead in their 
non-analytic behavior in $q$ since it is the one that will produce the logarithms. 

We also introduce the Sachs form factors:
\be
G_E(q^2)=F_1(q^2)+{q^2 \over 4m^2}F_2(q^2), \qquad G_M(q^2)=F_1(q^2)+F_2(q^2). 
\ee

We will also need the forward virtual-photon Compton tensor
\begin{equation}  \label{forw-amp2}
 T^{\mu\nu} = i\!\int\! d^4x\, e^{iq\cdot x}
  \langle {p,s}| {T J^\mu(x)J^\nu(0)} |{p,s}\rangle
\,,
\end{equation}
which has the following structure ($\rho=q\cdot p/m$):
\bea \label{inv-dec}
 T^{\mu\nu} =
  &\left( -g^{\mu\nu} + \frac{q^\mu q^\nu}{q^2}\right) S_1(\rho,q^2) \nn\\*
  & + \frac1{m_p^2} \left( p^\mu - \frac{m_p\rho}{q^2} q^\mu \right)
    \left( p^\nu - \frac{m_p\rho}{q^2} q^\nu \right) S_2(\rho,q^2) \nn \\*
  & - \frac i{m_p}\, \epsilon^{\mu\nu\rho\sigma} q_\rho s_\sigma A_1(\rho,q^2)
    \nn\\*
  & - \frac i{m_p^3}\, \epsilon^{\mu\nu\rho\sigma} q_\rho
   \bigl( (m_p\rho) s_\sigma - (q\cdot s) p_\sigma \bigr) A_2(\rho,q^2),
\eea
depending on four scalar functions. It is usual to consider the Born
approximation of these functions. They read
\bea
 S_1^{\rm Born}(\rho,q^2) & =
  -2F_1^2(q^2) - \frac {2(q^2)^2\,G_{\rm M}^2(q^2)}{(2m_p\rho)^2-(q^2)^2}, \\
 S_2^{\rm Born}(\rho,q^2) & =
  2\, \frac {4m_p^2q^2\,F_1^2(q^2)-(q^2)^2\,F_2^2(q^2)}{(2m_p\rho)^2-(q^2)^2}, \\
 A_1^{\rm Born}(\rho,q^2) & =  \label{Born.A1}
  -F_2^2(q^2) + \frac {4m_p^2q^2\,F_1(q^2)G_{\rm M}(q^2)}{(2m_p\rho)^2-(q^2)^2}, \\
 A_2^{\rm Born}(\rho,q^2) & =
  \frac {4m_p^3\rho\,F_2(q^2)G_{\rm M}(q^2)}{(2m_p\rho)^2-(q^2)^2}.
\eea

In the following section, we will use the results of Ji and Osborne \cite{JO}. Their 
notation relates to ours in the following way (for the spin-dependent terms):
$S_1^{JO}=A_1/m_p^2$ and $S_2^{JO}=A_2/m_p^3$. Note however that 
$A_1^{\rm Born}$ above is different from the $\bar S_1$
definition in Ref. \cite{JO} by the $F_2^2$ term.

\section{Matching}

The matching between HBET and NRQED can be performed in a generic
expansion in $1/m_p$, $1/m_\mu$ and $\al$.  We have two sort of loops:
chiral and electromagnetic.  The former are always associated to
$1/(4\pi F_0)^2$ factors, whereas the latter are always suppressed by
$\al$ factors. Any scale left to get the dimensions right scales with
$m_\pi$. In our case we are only concerned in obtaining the matching
coefficients of the lepton-baryon operators of NRQCD with
$O(\al^2\times(\ln m_q, \ln \Delta, \ln m_{l_i}))$
accuracy. Therefore, the piece of the Lagrangian we are interested in
reads
\be
\delta {\cal L} =
\displaystyle\frac{c_{3,NR}^{pl_i}}{m_p^2} N_p^{\dagger}
  N_p \ {l}^{\dagger}_i l_i
-\displaystyle\frac{c_{4,NR}^{pl_i}}{m_p^2} N_p^{\dagger}{\bfsigma}
  N_p \ {l}^{\dagger}_i{\bfsigma} l_i
  \,,
\ee
where we have not specified the lepton (either the electron or the muon).
In what follows, we will assume that we are doing the 
matching to NRQED($\mu$). Therefore, we have to keep the whole dependence on 
$m_{l_i}/m_\pi$. The NRQED($e$) case can then be derived by 
expanding $m_e$ versus $m_\pi$. In principle, a more systematic procedure 
would mean to go through QED($e$). Nevertheless, for this paper, as we do it 
will turn out to be the easier.

In principle, the contributions scaling with $1/m_\mu$ are the more
important ones. Nevertheless, they go beyond the aim of this paper.
This is specially so as far as we are only interested in logarithms and the
spin-dependent term $c^{pl_i}_{4,NR}$. Its general expression at $O(\al^2)$
reads (an infrared cutoff larger than $m_l\al$ is understood and the
expression for the integrand should be generalized for an eventual full
computation in $D$ dimensions)
\be
\label{c4}
c_{4,NR}^{pl_i}=-{i g^4 \over 3}\int {d^Dk \over (2\pi)^D}{1 \over k^2}{1 \over
k^4-4m_{l_i}^2k_0^2 }
\left\{
A_1(k_0,k^2)(k_0^2+2k^2)+3k^2{k_0 \over m_p}A_2(k_0,k^2)
\right\}
\,,
\ee
consistent with the expressions obtained long ago like in Ref. \cite{DS}. 
This expression has been obtained in the Feynman gauge. It correctly
incorporates the whole dependence on the lepton mass. Therefore, the
same expressions are valid for the hydrogen and muonic hydrogen.

In principle one should also consider contributions with one, two or
three electromagnetic current insertions in the hadronic matrix
elements instead of only two as in Eqs. (\ref{forw-amp2}) and
(\ref{c4}). Nevertheless, only the above contribute to the order of
interest.

Within the EFT framework the contribution from energies of $O(m_\rho)$
or higher in Eq. (\ref{c4}) are encoded in $c_{4,R}^{pl_i}\simeq
c_{4,R}^{p}$ (analogously for $c_3$). The contribution from energies
of $O(m_\pi)$ are usually split in three terms (actually this division
is usually made irrespectively of the energy which is being integrated
out): point-like, Zemach and polarizabilities corrections. They will
be discussed further later. From the point of view of chiral counting
the three terms are of the same order. Therefore, at the order of
interest we can divide $c_{4,NR}$ in the following way
\be
c_{4,NR}^{pl_i}=c_{4,R}^{p}+\delta c_{4,point-like}^{pl_i}
+\delta c_{4,Zemach}^{pl_i}+\delta c_{4,pol.}^{pl_i}
\,.
\ee 
Indeed, a similar splitting is usually done for $c_{3,NR}^{pl_i}$:
\be
c_{3,NR}^{pl_i}=
c_{3,R}^{p}
+\delta c_{3,point-like}^{pl_i}
+\delta c_{3,Zemach}^{pl_i}+\delta c_{3,pol.}^{pl_i} 
\ee

Let us stress at this point that we are only interested in the
logarithms. Therefore, we do not need to take care of the finite
pieces. This will significantly simplify the calculation.

\medskip

We obtain the following result for the point-like contribution
\be
\label{pointlike}
\delta c_{4,point-like}^{pl_i}={3+2c_F-c_F^2 \over 4}\alpha^2\ln{m_{l_i}^2
\over \nu^2} 
\,. 
\ee
Usually, in the literature, the computation of
this type of contributions is made considering the proton point-like
and relativistic, i.e. using standard QED-computations even at scales
of $O(m_p)$. The fact that
the proton has some anomalous magnetic moment that is due to hadronic
effects or, in other words, the fact that proton has structure makes
such theory no-renormalizable. This makes the result of the
computation divergent. Within this philosophy, the result to the hyperfine
splitting due to point-like contributions in Ref. \cite{DS} was
proportional to\footnote{Yet relativistic-type computations can be very
useful sometimes, like in identifying some $\ln(m_{l_i}/m_p) \rightarrow \ln(m_{l_i}/\nu)$
logarithms and $m_{l_i}/m_p$ corrections in an efficient way, see \cite{m/M}.} 
\be
{3+2c_F-c_F^2 \over 4}\alpha^2\ln{m_{l_i}^2
\over m_p^2}+{3 \over 4}(c_F-1)^2\ln{m_p^2 \over \Lambda^2}  
\,,
\ee  
where $\Lambda$ is the cutoff of this computation. This computation
would make sense if the scales on which the structure of the nucleon
appears were much larger than the mass of the nucleon (then $\Lambda$
could run up to this scale). Unfortunately, this is not the case and
the structure of the nucleon appears at scales of $O(m_p)$ or even
before. Therefore, to compute loops at the scale of $O(m_p)$ assuming
the proton point-like produces problems (divergences) as we have
seen. The procedure we use in this paper to deal with this issue is to
work with effective field theories where the nucleon is considered to
be non-relativistic. In other words, only at scales much smaller than
the mass of the nucleon it is a good approximation to consider the
nucleon point-like. Other usual method is to use some parameterization
of the form factors fitted to the experimental data (see for instance
\cite{BY,HFfitting}). This regulates the ultraviolet divergences
providing predictions for the hadronic correction to the hyperfine
splitting. This is a very reasonable attitude in the cases where we
are mainly interested in getting a number for the hadronic correction to
the hyperfine splitting. Nevertheless, this is not the procedure we
will follow in this paper, since our aim here is to gain as much
understanding as possible on the structure of the proton from QCD and
chiral symmetry. We want to understand how much of the coefficient
can be understood from logarithms and energies of $O(m_\pi)$, for which a
chiral Lagrangian can be used. This is something that could not be done
with the standard form factors used to fit the experimental data,
since they do not incorporate the correct momentum dependence at low
energies due to chiral symmetry.

\medskip

The Zemach correction is due to what is called the {\it elastic}
contribution, Eq. (27) of Ref. \cite{JO} for $A_1$ and analogously for
$A_2$ (nevertheless $A_2$ does not appear to give contribution). It
reads
\be
\label{Zemachc4}
\delta c_{4,Zemach}^{pl_i}= 
(4\pi\alpha)^2m_p{2 \over 3}\int {d^{D-1}k \over (2\pi )^{D-1}}
{1 \over {\bf k}^4}G_E^{(0)}G_M^{(2)}
\,.
\ee
Equation (\ref{Zemachc4}) can be obtained directly from Eq. (\ref{c4})
or working directly with the non-relativistic expressions and
introducing the form factors. Either way, it is comforting to find the
Zemach expression \cite{Zemach}.

The upper index in $G_E$ and $G_M$ has to do with the chiral
counting. $G_E^{(0)}=1$.  It is illustrative to split the contribution
to $G_M^{(2)}$ from $u$ and $d$, and the $\Delta$: 
$G_M^{(2)}=G_{M,u,d}^{(2)}+G_{M,\Delta}^{(2)}$ and analogously for the
Zemach contribution
\be
\delta c_{4,Zemach}^{pl_i}=\delta c_{4,Zemach-u,d}^{pl_i}+\delta
c_{4,Zemach-\Delta}^{pl_i}
\,.
\ee

Another strong simplification comes from the fact that we are just
searching for logarithms.  Therefore, we are only interested in the behavior
of the form factors for $m_p \gg k \gg m_\pi$ and not analytical in
${\bf k}^2$. In particular, for the logarithms, we are only interested in
the linear behavior in $|{\bf k}|$. From Ref. \cite{BKKM,BFHM} (see
also \cite{thesisHemmert}), we obtain
\be
G_{M,u,d}^{(2)}\doteq{m_p \over (4\pi F_0)^2}{\pi^2 \over 12}|{\bf k}|[-3g_A^2]
\,,
\ee
\be
G_{M,\Delta}^{(2)}\doteq{m_p \over (4\pi F_0)^2}{\pi^2 \over 12}|{\bf k}|
\left[{-4g_{\pi N \Delta}^2 \over 3}\right]
\,,
\ee
\be
\label{Zemachud}
\delta c_{4,Zemach-u,d}^{pl_i}
\simeq
{m_p^2 \over (4\pi F_0)^2}\alpha^2{2 \over 3}\pi^2g_A^2\ln{m_\pi^2 \over \nu^2}
\,,
\ee
\be
\delta c_{4,Zemach-\Delta}^{pl_i}
\label{ZemachDelta}
\simeq
{m_p^2 \over (4\pi F_0)^2}\alpha^2{8 \over 27}\pi^2g_{\pi N \Delta}^2
\ln{\Delta^2 \over \nu^2}
\,.
\ee
It is remarkable that the above results are $\pi$-enhanced.

Just for completeness, we also give the expression for 
$\delta c_{3,Zemach}^{pl_i}$:
\be
\delta c_{3,Zemach}^{pl_i}= 
4(4\pi\alpha)^2m_p^2m_{l_i}\int {d^{D-1}k \over (2\pi )^{D-1}}
{1 \over {\bf k}^6}G_E^{(0)}G_E^{(2)}
\,.
\ee
This term appears to be finite to the order of interest (it produces
no logarithms) and it agrees with the result obtained by Pachucki \cite{lambshift}
at leading order.

The Zemach corrections (both the spin-dependent and the
spin-independent) correctly incorporate the whole dependence on the
lepton mass. Therefore, the same expressions are valid for the
hydrogen and the muonic Hydrogen.

\medskip

Let us now consider the polarizability contributions. In the $SU(2)$
case (and including the $\Delta$) they should come from Eqs. 
(30,32,33,34,35,36) of Ref. \cite{JO}. In principle, in our case, one
should consider more diagrams besides those plotted in Ref.
\cite{JO}, like the ones due to the Wess-Zumino anomaly
action\footnote{We thank T. Hemmert for stressing this possibility to us.}. Nevertheless, as already stated in the introduction of the
pionic Lagrangian, it turns out that they do not contribute in
our case. Finally, we split the
polarizability contribution as follows (for the SU(2) case):
\be
\delta c_{4,pol.}^{pl_i}=\delta c_{4,pol.-\Delta}^{pl_i}+\delta
c_{4,pol.-\pi N}^{pl_i}+\delta c_{4,pol.-\pi\Delta}^{pl_i}
\,.
\ee
It is again a great simplification the fact that we are only searching
for logarithms. 

From Eqs. (32,35) of Ref. \cite{JO} we obtain (we also checked this
result by doing the computation directly in the non-relativistic limit)
\be
\label{polDelta}
\delta c_{4,pol.-\Delta}^{pl_i}={b_{1,F}^2 \over 18}\alpha^2\ln{\Delta^2
\over \nu^2}  
\,,
\ee 
where $b_1^F=G_1$ according to the definition in Ref. \cite{JO}. The
consequences this result has in a large $N_c$ analysis are
remarkable enough. In the large $N_c$, $b_1^F=3/(2\sqrt{2})\mu_V$
according to the Ji and Osborne definitions, where $\mu_V$ stands for the
isovector magnetic moment. On the other hand, by using the results of Ref.
\cite{KarlPaton} $\mu_p/\mu_n=-1$, one obtains (for practical purposes)
$\mu_p=\mu_V/2$ in the large $N_c$ . It follows that in this limit the
role of the delta is to cancel {\it all} the $\mu_p$ contribution in 
Eq. (\ref{pointlike}) ($(3+2c_F-c_F^2)/ 4=1-\mu_p^2/4$), which
effectively becomes the result of a point-like particle.

From Eqs. (30) and (34) of Ref. \cite{JO} we obtain
\be
\label{polpiN}
\delta c_{4,pol.-\pi N}^{pl_i}
=
-{m_p^2 \over (4\pi F_0)^2}g_{A}^2{\alpha^2 \over \pi}
{8\over 3}C\ln{m_\pi^2 \over \nu^2}
\,,
\ee
where $C$ is defined in the Appendix.

From Eqs. (33) and (36) of Ref. \cite{JO} we obtain
\be
\label{polpidelta}
\delta c_{4,pol.-\pi \Delta}^{pl_i}
=
{m_p^2 \over (4\pi F_0)^2}g_{\pi N\Delta}^2{\alpha^2 \over \pi}
{64\over 27}C\ln{\Delta^2 \over \nu^2}
\,.
\ee
It is worth noting that Eqs. (\ref{polpiN}) and (\ref{polpidelta})
cancel each other in the large $N_c$ limit, since $g_{\pi
N\Delta}=3/(2\sqrt{2})g_A$ in this case with the definitions of
Ref. \cite{JO}. Moreover, they are suppressed by $1/\pi$ factors and
the smallness of the numerical coefficient compared with the Zemach
term. 

Let us note that Eqs. (\ref{polDelta}), (\ref{polpiN}) and
(\ref{polpidelta}) may bring some light on why the polarization term
is much smaller than the Zemach term in a model-independent way since
we have an almost analytical result.

\medskip

Our results can be summarized in Eqs. (\ref{pointlike}),
(\ref{Zemachud}), (\ref{ZemachDelta}), (\ref{polDelta}),
(\ref{polpiN}) and (\ref{polpidelta}). The above computation has been
performed in SU(2), it would be interesting to repeat the analysis for
SU(3). Indeed, we can compute the Zemach correction due to the strange
quark (if we do not consider the spin 3/2 baryons) by using the
results of Ref. \cite{HKM} for $G_{M,s}$:
\be
G_{M,s}^{(2)}\doteq{m_p \over (4\pi F_0)^2}{\pi^2 \over 12}|{\bf
k}|[5D^2-6DF+9F^2]. 
\ee
We then obtain
\bea
\delta c_{4,Zemach-s}^{pl_i}&=& 
(4\pi\alpha)^2m_p{2 \over 3}\int {d^{D-1}k \over (2\pi )^{D-1}}
{1 \over {\bf k}^4}G_E^{(0)}G_{M,s}^{(2)}
\nn
\\
&\simeq&
-{m_p^2 \over (4\pi F_0)^2}\alpha^2{2 \over
9}\pi^2(5D^2-6DF+9F^2)\ln{m_K^2 \over \nu^2} 
\,.
\eea

In order to obtain a complete result, one should add the
strange-related contribution to the Zemach correction due to the
baryon spin 3/2 multiplet and to obtain the whole strange contribution
to the polarizability. This would require to have $A_1$ and $A_2$ for
SU(3) and including the baryon spin 3/2 multiplet, which are
unfortunately unknown. In any case, one may wonder whether, for the
polarizability corrections, the large
$N_c$ cancellation would also hold in this case as well as the $1/\pi$
and numerical factor suppression, so that it would be a very tiny
contribution as in the SU(2) case.

\subsection{Matching to pNRQED: Energy correction}

With the above results one can obtain the leading hadronic
contribution to the hyperfine splitting. It reads
\be
E_{\rm HF}=4{c^{pl_i}_{4,NR} \over m_p^2}{1 \over \pi} (\mu_{l_ip}\al)^3
\,.
\ee
By fixing the scale $\nu=m_\rho$ we obtain the following number for
the total sum in the SU(2) case:
\be
\label{HFtotal}
E_{\rm HF,logarithms}(m_\rho)=- 0.031\;{\rm MHz}
\,.
\ee 
The absolute value of this number would increase for a larger value of
$\nu$ and decrease for a smaller value. The main contribution to
Eq. (\ref{HFtotal}) comes from the Zemach and point-like corrections:
\bea
E_{\rm HF,Zemach-u,d}(m_\rho)&=&- 0.022\;{\rm MHz}
\,,
\\ 
E_{\rm HF,Zemach-\Delta}(m_\rho)&=&- 0.004\;{\rm MHz}
\,,
\\ 
E_{\rm HF,point-like}(m_\rho)&=&- 0.003\;{\rm MHz}
\,.
\eea 

Equation (\ref{HFtotal})
 accounts for approximately 2/3 of the difference between theory (pure QED)
\cite{BY} and experiment \cite{exp}:
\be
E_{\rm HF}(QED)-E_{\rm HF}(exp)=- 0.046\;{\rm MHz}.
\ee 
What is left gives the expected size of the counterterm.
Experimentally what we have is $c_{4,NR}=-48\alpha^2$ and
$c_{4,R}(m_\rho)\simeq -16\alpha^2$. This last figure gives the
 expected size
of the counterterm of the Lagrangian. A more detailed analysis would
require to work in an specific scheme (for instance MS) to fix the
finite pieces. We expect to come back to this issue in the future. A
point to stress is that this number is universal, i.e. the same for the
electron and for the muon up to corrections suppressed by the ratio
of the lepton mass versus the proton mass:
\be
c_{4,R}^{pe}(m_\rho)\simeq c_{4,R}^{p\mu} (m_\rho)
\,.
\ee
This observation could be used in an eventual measurement of the 
hyperfine splitting of the muonic hydrogen.

\medskip

The introduction of the partial SU(3) computation would worsen the
above prediction by  
\be
E_{\rm HF,Zemach-kaon}(m_\rho)= 0.003\;{\rm MHz}
\ee 
bringing the total sum down to $-0.027\;{\rm MHz}$ and
$c_{4,R}(m_\rho)$ to $ -20\alpha^2$. 

\section{Conclusions}

We have performed a first exploratory study on the application of
effective field theories emanated from chiral Lagrangians on atomic
physics\footnote{The pionium, which has received quite attention
recently \cite{pionium1,pionium2,pionium3,pionium4,pionium5,pionium6},
has also been studied within a similar non-relativistic effective
field theory philosophy
\cite{pionium1,pionium3,pionium4,pionium6}. Specially close to ours is
the approach followed in \cite{pionium4}. The pionic hydrogen
has also been studied using effective field theories very recently \cite{pionichyd}.}.  We have computed the
$c^{pl_i}_{4,NR}$ matching coefficient of the NRQED Lagrangian for the
$e$-$p$ and $\mu$-$p$ sector with $O(\al^2\times(\ln m_q,\ln \Delta,
\ln m_{l_i}))$ accuracy.  The Hyperfine splitting of the hydrogen and
muonic hydrogen has been computed with
$O(m_{l_i}^3\al^5/m_p^2\times(\ln m_q,\ln \Delta, \ln m_{l_i}))$
accuracy. We note that our results include the complete expression for
the leading chiral logarithms.
  
The difference between the experimental value of the hydrogen
hyperfine splitting and the pure QED computation reads
\be
E_{\rm HF}(QED)-E_{\rm HF}(exp)=- 0.046\;{\rm MHz},
\ee 
whereas our theoretical prediction reads
\be
\label{totallogs}
E_{\rm HF,logarithms}(m_\rho)=- 0.031\;{\rm MHz}
\,.
\ee 
We are then able to obtain an estimate of $c^p_{4,R}(m_\rho)\simeq
-16\alpha^2$ which is valid for both $e$-$p$ and $\mu$-$p$
systems. 

One could improve these results by performing the whole computation in
the MS scheme or alike (in this case some of the expressions in this
paper should be rewritten in D-dimensions). It would follow that not
only the logarithms but the finite pieces (in a specific scheme) of the
matching coefficient would be obtained too. This would fix with a
greater precision the value of $c^p_{4,R}(m_\rho)$ and, thus, the
respective size of the effects due to the physics at scales of
$O(m_\rho)$ and at scales of $O(m_\pi)$. This is important since the
experimental number is precise enough to give an accurate number for
$c^p_{4,R}(m_\rho)$, which could be used to test models at $O(m_\rho)$
scales. To perform the full computation in SU(3) would be also highly
desirable. A partial SU(3) result brings Eq. (\ref{totallogs}) down to
$- 0.027\;{\rm MHz}$.

Our results may help to better understand the fact that the Zemach
correction is much larger than the polarizability contribution, since
we have (almost) analytical expressions for these contributions. The
polarizability term (except Eq. (\ref{polDelta})) vanishes in the
large $N_c$ and it is $1/\pi$ and numerical-factor suppressed with
respect the Zemach terms. On the other hand, Eq. (\ref{polDelta}), in
the large $N_c$ limit, cancels {\it all} the $\mu_p$ contribution in
Eq. (\ref{pointlike}), which effectively becomes the result of a
point-like particle.

Several lines of research are worth pursuing. One is trying to compute
$c_{3,NR}$ within HBET, since its numerical value could be obtained
from measurements of the Lamb shift, and it is related with (and in a
way defines) the proton radius. Another could be to consider more
complicated atoms within this effective field theory formalism (see,
for instance \cite{Yelkhovsky} for the Helium).

\medskip

{\bf Acknowledgments} \\ 
We thank T. Hemmert and J. Soto for useful discussions and X.-D. Ji
for a correspondence. We also thank J. Soto for the reading of the 
manuscript. This work is supported by MCyT and Feder
(Spain), FPA2001-3598, and by CIRIT (Catalonia), 2001SGR-00065.

\appendix

\Appendix{Constants}
\bea
F_0&=&92.5\,{\rm MeV}\,,
\\
g_A&=&1.25\,,
\nn
\\
m_\pi&=&140 \,{\rm MeV}\,,
\nn
\\
m_p&=&m_n=938 \,{\rm MeV}\,,
\nn
\\
\Delta&=&294\,{\rm MeV}\,,
\nn
\\
g_{\pi N\Delta}&=&1.05\,,
\nn
\\
b_1^F&=&3.86\,,
\nn
\\
\nn
F&=&1/2\,,
\\
\nn
D&=&3/4\,,
\\
\nn
m_\rho&=&770 \,{\rm MeV}\,.
\eea
$b_1^F$ and $g_{\pi N\Delta}$ have been obtained from the decays of the
delta in the non-relativistic limit (consistent with the accuracy
of our calculation).
 
The values of $F$ and $D$ are consistent with the large $N_c$
limit. Finally 

\bea
C&=&2\,\int _{0}^{1}\int _{0}^{1}
       {\sqrt{1 - y^2}}\,
         \left( -2\,x\,\left( 2 + y^2 \right)  + {1 \over y}
           \left( 2\,\left( 1 - x \right) \,x\,
                 \left( 2 + y^2 \right) \,
                 {\sqrt{\frac{1}{x - x^2 + x^2\,y^2}}} 
\right.
\right.
\nn
\\
&&
\left.
\left.
- 
                3\,\left( 1 - 2\,x \right) \,y^2\,
                 {\sqrt{\frac{x}
                     {1 - x\,\left( 1 - y^2 \right) }}}
                \right) \,
              {\rm Sinh}^{-1}({\sqrt{\left( \frac{x}
                     {1 - x} \right) }}\,y) \right) \,d
        y\,dx
\nn
\\
&=&-0.165037
\,.
\eea

\end{document}